\documentclass[aps,pre,preprint,groupedaddress,floatfix,longbibliography]{revtex4-1}

\usepackage{amsmath}
\usepackage{graphicx}
\usepackage{color}

\begin{document}
\title{Plasmonic metagratings for simultaneous determination of Stokes parameters}

\author{Anders Pors}
\email[]{alp@iti.sdu.dk}
\affiliation{Department of Technology and Innovation, University of Southern Denmark, Niels Bohrs All{\'{e}} 1, DK-5230 Odense M, Denmark}

\author{Michael G. Nielsen}
\affiliation{Department of Technology and Innovation, University of Southern Denmark, Niels Bohrs All{\'{e}} 1, DK-5230 Odense M, Denmark}

\author{Sergey I. Bozhevolnyi}
\affiliation{Department of Technology and Innovation, University of Southern Denmark, Niels Bohrs All{\'{e}} 1, DK-5230 Odense M, Denmark}

\begin{abstract}
Measuring light's state of polarization is an inherently difficult problem, since the phase information between orthogonal polarization states is typically lost in the detection process.
In this work, we bring to the fore the equivalence between normalized Stokes parameters and diffraction contrasts in appropriately designed phase-gradient birefringent metasurfaces and introduce a concept of all-polarization birefringent metagratings. The metagrating, which consists of three interweaved metasurfaces, allows one to easily analyze an arbitrary state of light polarization by conducting simultaneous (i.e., parallel) measurements of the correspondent diffraction intensities that reveal immediately the Stokes parameters of the polarization state under examination. Based on plasmonic metasurfaces operating in reflection at the wavelength of 800\,nm, we design and realize phase-gradient birefringent metasurfaces and the correspondent metagrating, while experimental characterization of the fabricated components convincingly demonstrates the expected functionalities. We foresee the use of the metagrating in compact polarimetric setups at any frequency regime of interest.
\end{abstract}



\maketitle

\section{Introduction}
Monochromatic electromagnetic waves are typically characterized by their intensity, frequency, and the state of polarization, with the first two characteristics being easy to assess using suitable detectors and spectrometers. The state of polarization, on the other hand, is an inherently tricky problem to experimentally probe, since the phase information between orthogonal polarization states is completely lost in the conventional detection process (sensitive only to the light intensity or power). As such, the determination of the polarization typically necessitates a series of measurements with properly arranged polarizers consecutively placed in front of the detector, allowing one to eventually obtain the Stokes parameters that, similar to the ellipsometric parameters, fully describe the state of polarization\cite{bohren}. As a way to measure the state of polarization in one shot, it is possible to parallelize the measurement process by splitting the incident beam into several beams and using multiple polarizers and detectors\cite{compain_1998}, although such an approach increases the size and complexity of the optical system. Despite the inconvenience in determining the polarization, it is in most applications of paramount importance to know this parameter since light-matter interactions are, in general, polarization-dependent. As archetypical examples, we mention plane wave reflection and transmission at material interfaces in which the Fresnel coefficients are different for orthogonal polarizations, and antennas that dominantly emit (or respond to) a specific polarization component. Moreover, polarimetry, which represents a branch of science that extracts information from the measurement of a probe signal's resulting polarization state, covers as diverse areas as thin film characterization\cite{rzhanov1979} (i.e., ellipsometry), astronomy\cite{tinbergen_1996}, and biology\cite{collings_1997}. Consequently, it is clear that a simple, fast, and compact way of measuring the state of polarization is a desirable feature in a broad range of applications. In the reverse situation of synthesizing an arbitrary polarization, recent progress has been conducted at the telecom wavelength by utilizing silicon-on-insulator waveguides combined with an out-coupling nanoantenna\cite{fortuno_14}, while suggestions of extending the principle to arbitrary polarization detection have also been expressed\cite{carbonell_2014,fortuno_2014_2}.

In the context of polarization-dependent light-matter interaction, it is appropriate to discuss the recent advances with carefully designed plasmonic (i.e., metallic) metasurfaces. Metasurfaces are characterized by a subwavelength thickness in the direction of beam propagation, while the transverse plane consists of an array of metallic particles with subwavelength periodicity\cite{yu,yu3}. Accordingly, incident light does not resolve the fine-structure of the metasurface and it can, for this reason, be considered as an effective interface discontinuity that may control both the amplitude and phase of the reflected and/or transmitted light. Particularly, early work has considered transmission through a space-variant metal-stripe metasurface\cite{gori_99,bomzon_01}, which results in part of the incident light being redirected (i.e., diffracted), similar to the functionality of blazed gratings\cite{larouche_12}. Moreover, the response of the metasurface is partially polarization-dependent and, consequently, it can be used as part of an optical system for probing the state of polarization of the incident light. For example, by combining the metasurface with a linear polarizer and performing two successive measurements\cite{gori_99,wen_2015} or, alternatively, by Fourier transforming the transmitted near-field when a quater-wave plate is placed in front of the metasurface \cite{bomzon_01}, it is possible to determine the Stokes parameters.

In recent years, considerably amount of work has been devoted to realizing planar optical components, like wave plates\cite{hao,pors_2011,khoo_2011,zhao2,roberts_2012}, lenses\cite{verslegers,lin,ishii,aieta2,chen_2012} and gratings\cite{yu,ni,sun2}, that typically either manipulate the state of polarization, propagation direction, or wave front of the incident light. For efficient manipulation of light, however, it is advantageous to consider configurations based on the low-frequency concept of reflectarrays\cite{pozar_1997}, also known in the optical regime as gap-surface plasmon (GSP) based metasurfaces\cite{pors2,pors3} due to the physical origin of the plasmonic resonances\cite{bozhevolnyi_2007}. These metasurfaces consist of an optically-thick metal film overlaid by a nanometer-thin dielectric spacer and an array of space-variant metal nanobricks (sometimes also denoted nanopatches). The gradient metasurface only operates in reflection, but it has the advantage of being simple to fabricate (i.e., one-step of electron beam lithography) and (in the limit of negligible Ohmic losses) the potential to manipulate light with 100\% efficiency. Here, it is important to stress that the high efficiency is not only limited to linear polarization states\cite{sun2,pors2}, but also includes circularly polarized light\cite{zheng_2015,luo_2015}. The design principles are, however, slightly different. For linearly polarized light, the  $\sim2\pi$ phase control of the reflected light is reached by choosing (at the design wavelength) nanobrick dimensions at and near the fundamental GSP resonance, while appropriately selecting other geometrical parameters so that the metasurface remains highly reflective even at the resonance\cite{pors6}. For circularly polarized light, on the other hand, the phase gradient is geometrically induced by spatially varying the orientation of the identical nanobricks, hereby leading to a so-called Pancharatnam-Berry phase\cite{pancharatnam_1956,berry_1984}. Furthermore, it is important to notice that circularly polarized light of opposite handedness will by symmetry considerations always experience geometrically-induced phase gradients of opposite sign, meaning that the two orthogonal polarization states are split when reflected by the metasurface\cite{wen_2015}. In contrast, birefringent metasurfaces manipulating linear polarization states must be carefully designed to obtain a similar polarization-splitting functionality\cite{farahani,pors3}.

In this work, we first highlight the equivalence between the Stokes parameters and diffraction contrasts in three types of phase-gradient birefringent metasurfaces, where each metasurface functions as a polarization-splitter for a certain polarization basis. GSP-based metasurfaces are then designed at a wavelength of 800\,nm, while their functionalities are verified both numerically and experimentally. Finally, we interweave the three metasurfaces into a so-called metagrating that allow us, without the need of any additional polarizers, to determine the state of polarization of the incident light by simultaneously measuring the associated six diffraction intensities. As such, this work proposes a new compact optical component that allows for fast and simple determination of the state of polarization.

\section{Results}
\subsection{Stokes parameters}
Let us start by considering a monochromatic plane wave propagating along the $z$-direction with the amplitude of the electric field defined (in Cartesian coordinates) by the Jones vector
\begin{equation}
\mathbf{E}_0=
\begin{pmatrix} 
A_x \\
A_ye^{i\delta} 
\end{pmatrix},
\label{eq:E0}
\end{equation}
where $A_x$ and $A_y$ are real-valued positive constants and $\delta$ is the phase difference between the two components. It should be noted that $\mathbf{E}_0$ contains all the information about the intensity and polarization of the plane wave and, hence, constitutes a crucial parameter in any optical system. Nevertheless, the fact that optoelectronic detectors only measure intensities complicates the experimental determination of $\mathbf{E}_0$ since $\delta$ cannot be directly measured. 
Instead, one can make a set of measurements in which light transverses specifically oriented polarizers before reaching the detector, hereby allowing one to determine the Stokes parameters
\begin{align}
& s_0=A_x^2+A_y^2\propto I,\\
& s_1=A_x^2-A_y^2, \label{eq:s1}\\
& s_2=2A_xA_y\cos\delta= A_a^2-A_b^2, \label{eq:s2}\\
& s_3=2A_xA_y\sin\delta= A_r^2-A_l^2. \label{eq:s3}
\end{align}
It is clear that $s_0$ is proportional to the intensity, $I$, of the beam, whereas $s_1$-$s_3$ describe the state of polarization. Importantly, the latter three Stokes parameters can be found by measuring the respective intensities of the two components in the orthonormal bases $(\hat{\mathbf{x}},\hat{\mathbf{y}})$, $(\hat{\mathbf{a}},\hat{\mathbf{b}})=\tfrac{1}{\sqrt{2}}(\hat{\mathbf{x}}+\hat{\mathbf{y}},-\hat{\mathbf{x}}+\hat{\mathbf{y}})$, and $(\hat{\mathbf{r}},\hat{\mathbf{l}})=\tfrac{1}{\sqrt{2}}(\hat{\mathbf{x}}+i\hat{\mathbf{y}},\hat{\mathbf{x}}-i\hat{\mathbf{y}})$. Here, $(\hat{\mathbf{a}},\hat{\mathbf{b}})$ corresponds to a rotation of the Cartesian coordinate system $(\hat{\mathbf{x}},\hat{\mathbf{y}})$ by $45^\circ$ with respect to the $x$-axis, while $(\hat{\mathbf{r}},\hat{\mathbf{l}})$ is the basis for right- and left-circularly polarized light. Note that $s_1-s_3$ normalized by $s_0$ attain the maximum values of $\pm1$ for light being polarized in accordance with the respective coordinate axes. 
%
\begin{figure}
\centering
\includegraphics[width=7cm]{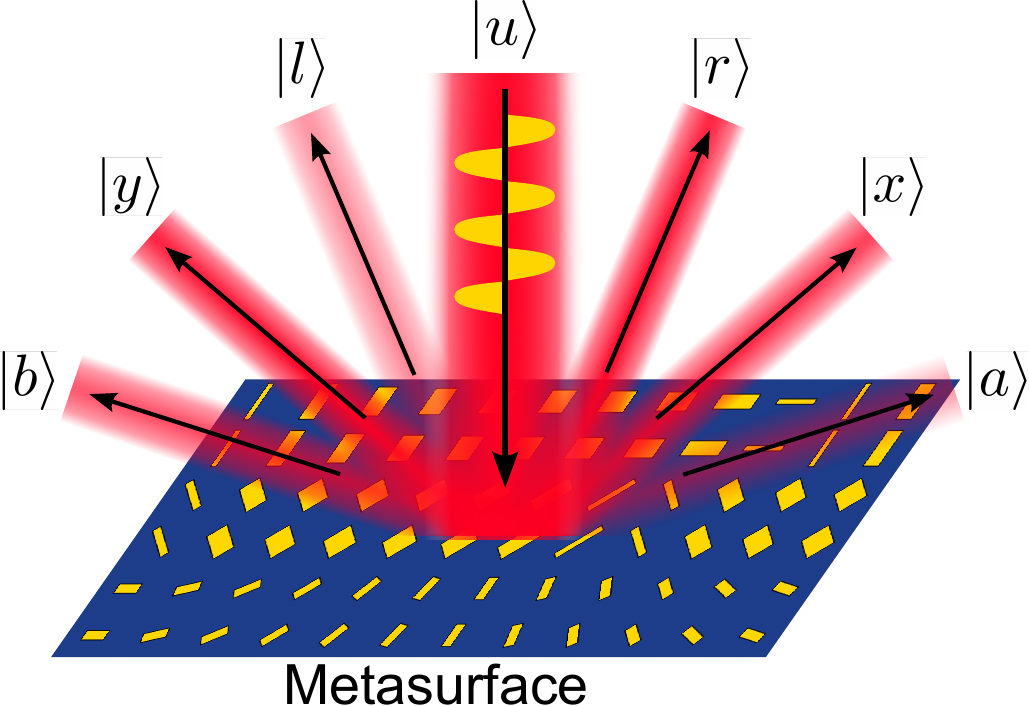}
\caption{Illustration of metagrating's working principle. An incoming beam with unknown polarization state $|u\rangle$ is on reflection being represented in the bases $(\hat{\mathbf{x}},\hat{\mathbf{y}})$, $(\hat{\mathbf{a}},\hat{\mathbf{b}})$, and $(\hat{\mathbf{r}},\hat{\mathbf{l}})$, hereby allowing one to determine the Stokes parameters and, hence, the state of polarization without the need of multiple measurements or an interferometric setup.\label{fig:sketch}}
\end{figure}
From the above review of Stokes parameters, it is clear that a fast and simple determination of the state of polarization requires an optical element that upon interaction with the light represents \textit{at once} the light in the three bases and splits the associated two orthogonal states. In this work, we realize such a functionality by designing a phase-gradient birefringent metagrating in reflection -- the working principle is illustrated in Fig. \ref{fig:sketch} where the incident beam is reflected into six diffraction orders each representing one of the fundamental states involved in the determination of the Stokes parameters [see Eqs. (\ref{eq:s1})-(\ref{eq:s3})].
%
\begin{figure*}[ht]
\centering
\includegraphics[width=16cm]{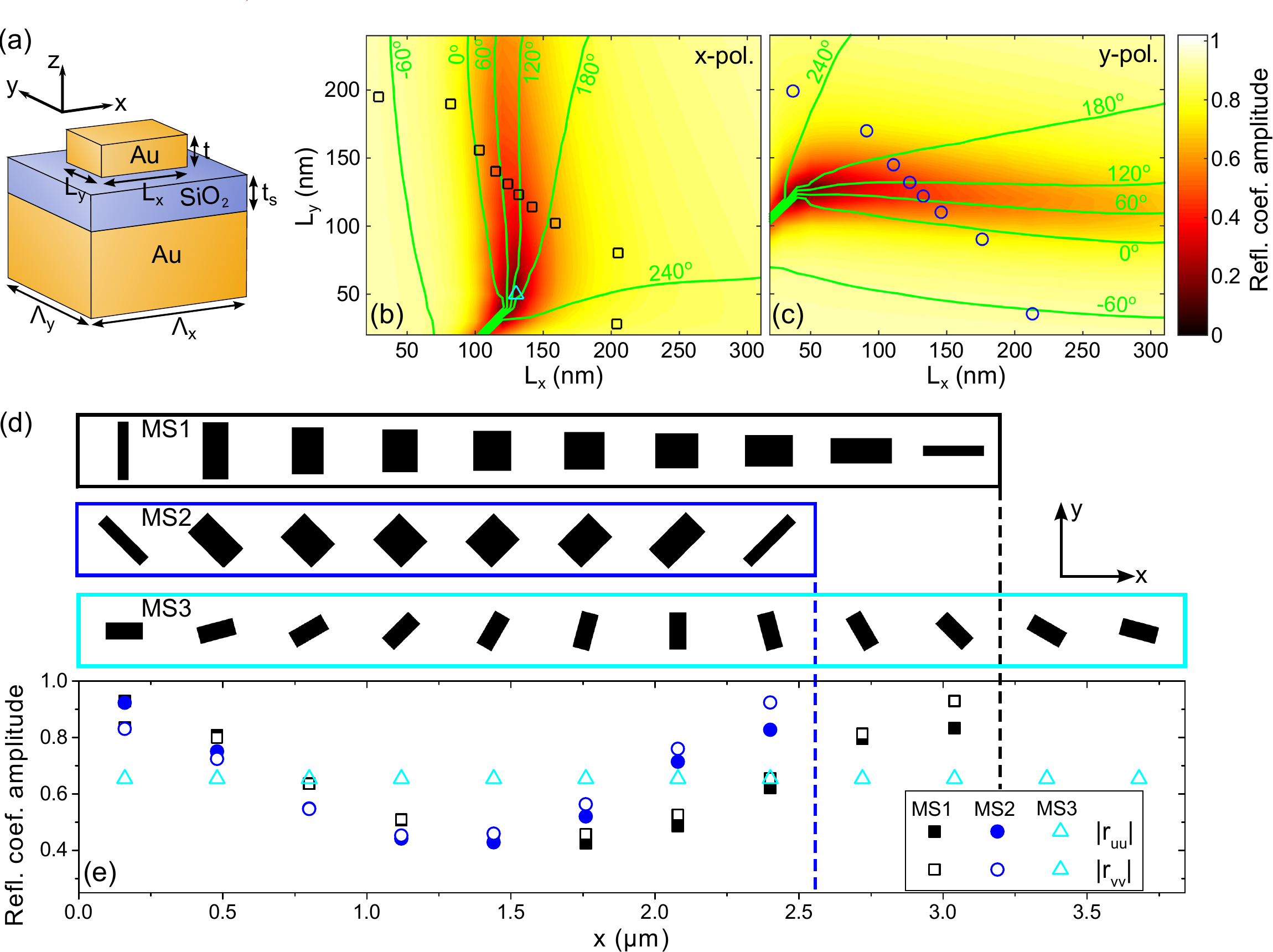}
\caption{Design of birefringent phase-gradient metasurfaces. (a) Sketch of basic unit cell consisting of gold nanobrick on top of a glass spacer and gold substrate. (b,c) Calculated reflection coefficient as a function of nanobrick widths for $x$- and $y$-polarized light, respectively, and geometrical parameters $\Lambda_x=320$\,nm, $\Lambda_y=250$\,nm, $t=40$\,nm, $t_s=50$\,nm, and wavelength $\lambda=800$\,nm. Color maps show the reflection coefficient amplitude, while green lines are contours of the corresponding reflection phase. Square, circle, and triangular markers, respectively, indicate the nanobrick sizes constituting the metasurface super cells depicted in (d). (e) Reflection coefficient amplitude along the three metasurfaces, as extracted from panel b,c, for polarization bases $(\hat{\mathbf{x}},\hat{\mathbf{y}})$, $(\hat{\mathbf{a}},\hat{\mathbf{b}})$, and $(\hat{\mathbf{r}},\hat{\mathbf{l}})$, respectively.    
\label{fig:design}}
\end{figure*}

\subsection{Phase-gradient birefringent metasurfaces}
Having defined the required functionality of an all-polarization sensitive metagrating, we now proceed with the discussion of phase-gradient birefringent metasurfaces that function as blazed gratings but possess phase-gradients of opposite signs for orthogonal polarizations, thus resulting in a splitting of the two polarization states. In this context, the ideal reflection coefficient is of the form
\begin{equation}
\bar{\bar{r}}=r\begin{pmatrix} 
 e^{i2\pi x/\Lambda} & 0\\
0 &  e^{-i2\pi x/\Lambda} 
\end{pmatrix},
\label{eq:r}
\end{equation}
where $r\leq 1$ is a real-valued positive constant, $\Lambda$ is the period of the grating, and $x$ is the spatial coordinate along the direction of phase variation. It should be noted that we have in previous work\cite{pors3} managed to implement the above reflection coefficient for which $x$- and $y$-polarized light experience phase gradients of opposite signs, meaning that the two orthogonal polarizations are split into $\pm 1$ diffraction order, respectively. We now make the crucial point that splitting of orthogonal polarizations is not limited to the $(\hat{\mathbf{x}},\hat{\mathbf{y}})$ basis, but the implementation of the \textit{diagonal} reflection matrix in \eqref{eq:r} for \textit{any} basis results in diffractive splitting of the associated orthogonal polarizations. A simple spatial Fourier transform of the reflected field in the basis ($\hat{\mathbf{u}},\hat{\mathbf{v}}$) shows that the intensities in the $\pm 1$ diffraction orders are
$$
I_{+1}\propto r^2A_u^2 \quad,\quad I_{-1}\propto r^2A_v^2,
$$
with the diffraction contrast defined by
\begin{equation}
D=\frac{I_{+1}-I_{-1}}{I_{+1}+I_{-1}}=\frac{A_u^2-A_v^2}{A_u^2+A_v^2}.
\label{eq:D}
\end{equation}
It is evident that the diffraction contrast is \textit{equal} to $s_i/s_0$ ($i=1,2,3$) for the bases $(\hat{\mathbf{x}},\hat{\mathbf{y}})$, $(\hat{\mathbf{a}},\hat{\mathbf{b}})$, and $(\hat{\mathbf{r}},\hat{\mathbf{l}})$, respectively. As such, we can construct an all-polarization sensitive metagrating that consists of three parts, each implementing \eqref{eq:r} for one of the three bases, while the associated diffraction contrasts correspond to the normalized Stokes parameters.

\subsection{Design of metasurfaces}
In the design of the birefringent GSP-based metasurfaces, we first consider a gold-glass-gold configuration with no phase gradient and light incident normally to the surface [Fig. \ref{fig:design}(a)]. 
Here, the nanobrick dimensions ($L_x$ and $L_y$) represent the only two variable parameters, while the remaining geometrical parameters are constant. Note that the larger dimension of the unit cell along the $x$-direction ($\Lambda_x=320$\,nm vs. $\Lambda_y=250$\,nm) is chosen solely for ensuring low diffraction angles (in the $xz$-plane) in the final metagrating design, thus making it suitable for optical characterization. Using the commercially available finite-element software Comsol Multiphysics (ver. 5.0), with the permittivity of gold described by interpolated experimental values\cite{johnson} while glass, assuming to be silicon dioxide, takes on the constant refractive index 1.45, we compute the complex reflection coefficient for $x$- and $y$-polarized light [Figs. \ref{fig:design}(b) and \ref{fig:design}(c)]. 
As seen, we find the metasurface to be highly reflecting for most nanobrick dimensions, only featuring a dip in reflection at the GSP resonance that, simultaneously, implies an almost $2\pi$ change in the reflection phase. It is this strong variation of the phase, together with the assumption of local periodicity, that allow us to design phase-gradient metasurfaces. The latter prerequisite refers to the assumption that the coupling between neighboring nanobricks is weak, so that the response of the individual elements is not noticeably affected by the exact shape and size of the nearby elements. 

In the design of polarization-splitter for the $(\hat{\mathbf{x}},\hat{\mathbf{y}})$ basis, which we denote metasurface 1 (MS1), we choose to discretize the phase space in 10 steps so that the super cell size (in the diffraction plane) is $10\Lambda_x$, resulting in incident light being diffracted into the angle $\pm 14.5^\circ$ with respect to the surface normal for $x$- and $y$-polarized light, respectively. The appropriate nanobrick dimensions are marked with squares in Fig. \ref{fig:design}(b), and the super cell is constructed by sequentially placing these elements with a center-to-center separation of $\Lambda_x$ [Fig. \ref{fig:design}(d), black contour]. It should be noted that controlling the reflection phase along the super cell for both polarizations by only two geometrical parameters entails no direct way of regulating the associated reflection amplitudes. Ideally, cf. \eqref{eq:r}, the reflection amplitudes should be constant along the metasurface, but such a criteria cannot be perfectly fulfilled with the rather simple unit cell employed in this work, since nanobrick elements close to the GSP resonance will experience more pronounced absorption losses, hereby making the reflection amplitudes position-dependent [Fig. \ref{fig:design}(e)].

The second phase-gradient metasurface (MS2), which should split light of orthogonal polarizations in the basis $(\hat{\mathbf{a}},\hat{\mathbf{b}})$, can in fact be designed in a completely equivalent way as described for MS1, just with the nanobrick elements rotated $45^\circ$ with respect to the $x$-axis. In order to ensure that diffraction spots from different metasurfaces do not spatially overlap, we here choose to discretize the reflection phase in 8 steps, leading to diffraction angles of $\pm 18.2^\circ$. The appropriate nanobrick dimensions are marked with circles in Fig. \ref{fig:design}(c), with a top-view of the super cell and the related reflection amplitude variation shown in Figs. \ref{fig:design}(d) and \ref{fig:design}(e), respectively. Finally, the third metasurface (MS3), which is intended to separate circularly polarized light of opposite handedness, is constructed by implementing a geometric phase covering all $2\pi$ in 12 elements [Fig. \ref{fig:design}(d), cyan contour], hereby resulting in first order diffraction angles of $\pm 12.0^\circ$. In order to suppress zero-order (i.e., specular) diffraction for the benefit of first-order diffraction in MS3, the dimension of the nanobrick [marked by a triangle in Fig. \ref{fig:design}(b)] is chosen so that only light polarized along the short-axis is efficiently reflected which, at the same time, also implies a phase difference between light polarized along the short- and long-axis of the nanobrick of $\sim \pi$. This design strategy, of course, entails noticeable absorption losses, but the time-averaged reflection for circularly polarized light is still on the level of the other metasurfaces, as shown in Fig. \ref{fig:design}(e). We recognize that the efficiency of MS3 can be further optimized by utilizing a nanobrick unit cell with properties akin to a half-wave plate, as demonstrated in other studies\cite{bomzon_2002,kang_2012,zheng_2015,luo_2015}.
%
\begin{figure}[t]
\centering
\includegraphics[width=8.6cm]{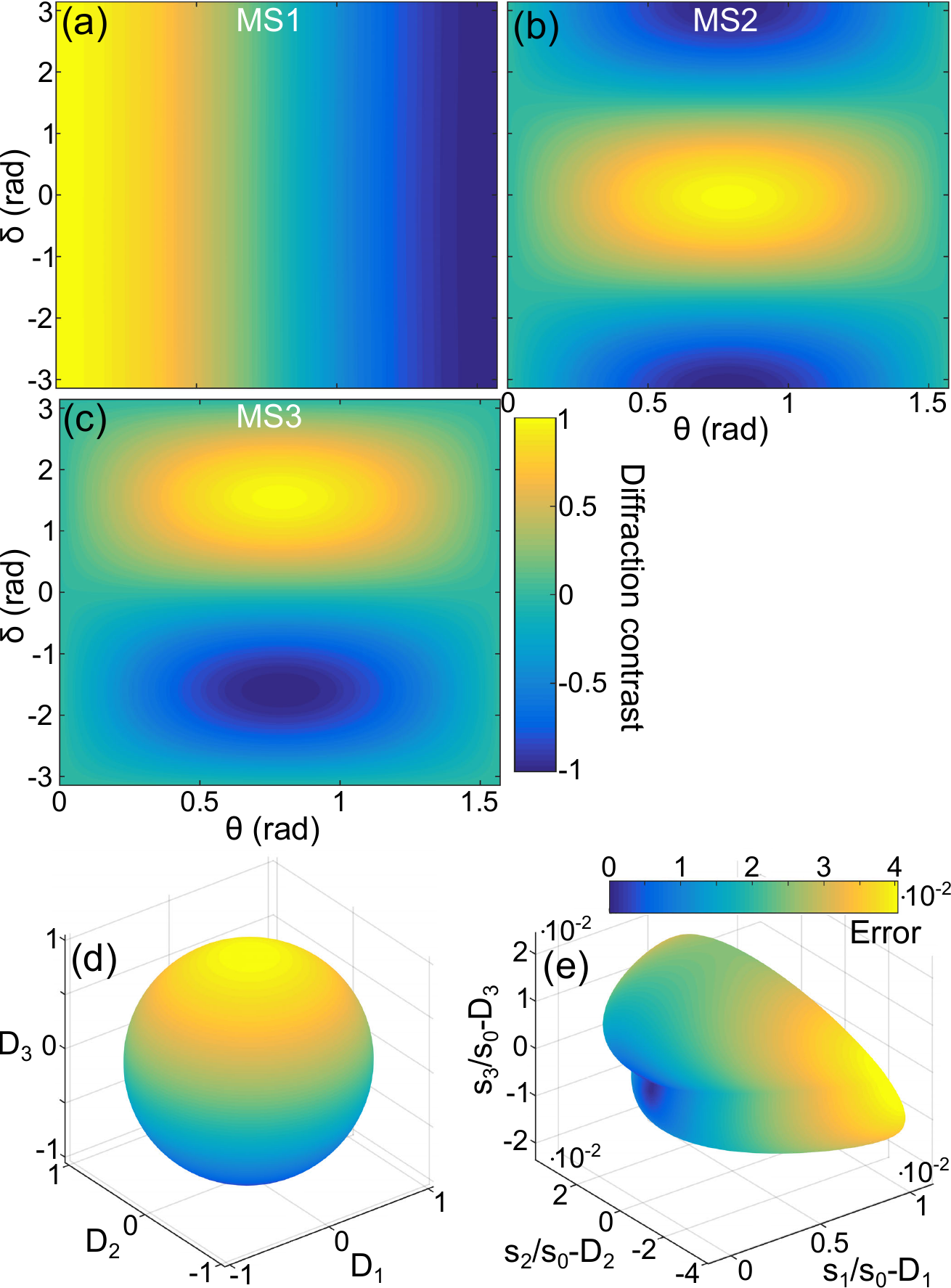}
\caption{Theoretical performance of phase-gradient metasurfaces. (a-c) Diffraction contrasts for the three metasurfaces as a function of polarization state of incident light. Here, the amplitude of the electric field is defined as $(A_x,A_y)=(\cos\theta,\sin\theta)$ and $\delta$ is the phase difference. (d) Three-dimensional plot of the diffraction contrasts, clearly illustrating that the associated surface closely resembles a unit sphere. (e) Plot of the difference between calculated diffraction contrasts and the Stokes parameters, with the colour bar representing the 2-norm error $\left\|\mathbf{S}-\mathbf{D}\right\|_2$ [Here, $\mathbf{S}=(s_1,s_2,s_3)/s_0$ and $\mathbf{D}=(D_1,D_2,D_3)$]. \label{fig:theory}}
\end{figure}

\subsection{Theoretical performance}
It is evident from the above discussion that a successful design of phase-gradient metasurfaces relies on several approximations, such as discretization of the phase gradient, local periodicity assumption, and non-uniform reflection amplitude (not an issue for MS3). In order to gauge how these non-idealities affect the performance of the metasurfaces, we have conducted full-wave numerical simulations of each of the metasurfaces, calculating for all possible polarization states the associated diffractions contrasts [Figs. \ref{fig:theory}(a)-\ref{fig:theory}(c)]. 
%
%
\begin{figure*}[t]
\centering
\includegraphics[width=14cm]{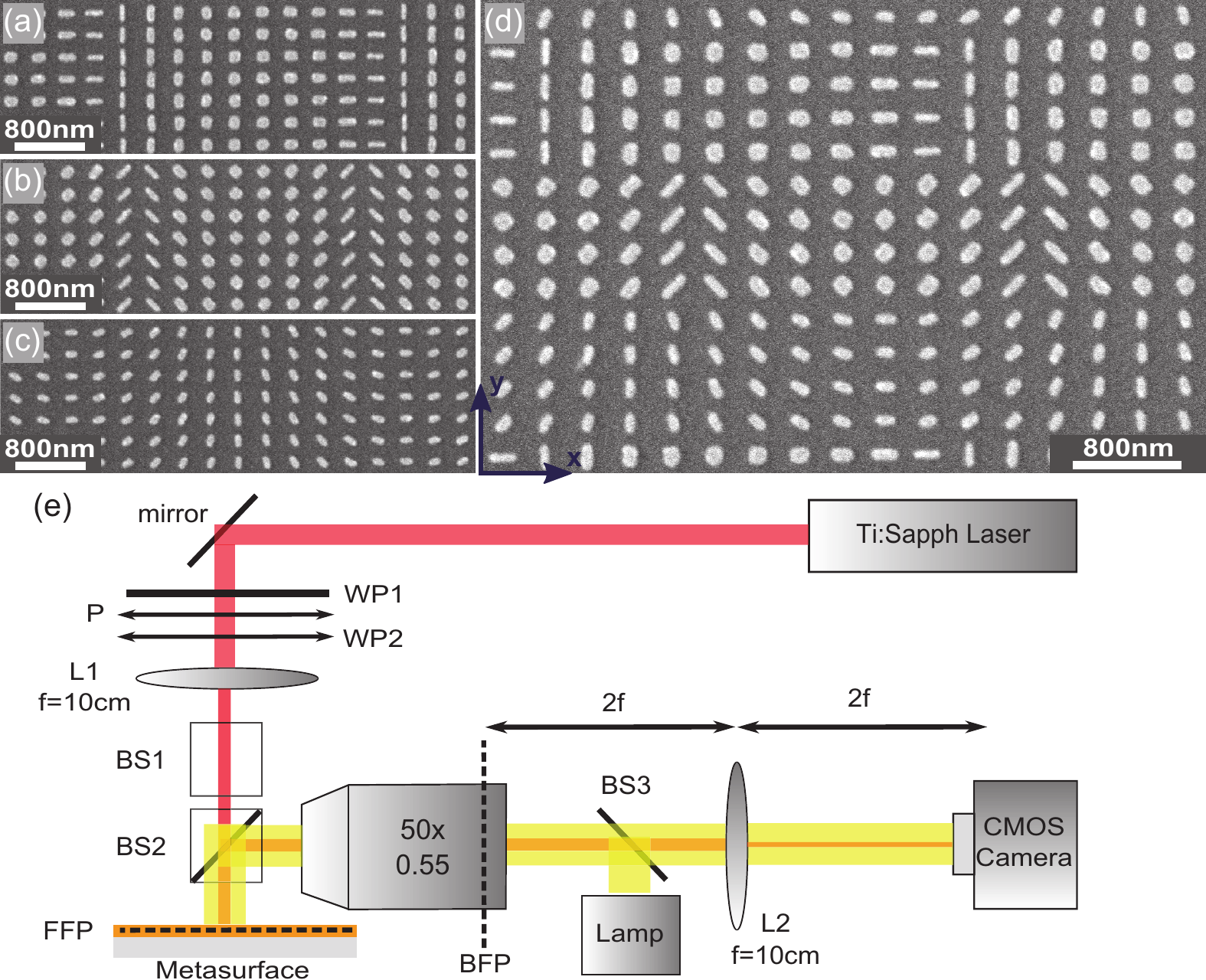}
\caption{Fabricated metasurfaces and experimental setup. Representative scanning electron microscopy images of (a-c) metasurface 1-3, respectively, and (d) metagrating. (e) Schematic of the experimental setup for optical characterization.    
\label{fig:exp1}}
\end{figure*}
The diffraction contrasts are presented as a function of the phase difference $\delta$, as defined in \eqref{eq:E0}, and the angle $\theta$, which is related to the electric field components by $(A_x,A_y)=(\cos\theta,\sin\theta)$. As requested, MS1 displays a diffraction contrast that only depends on $\theta$, with maximum values of $\pm1$ reached for purely $x$- and $y$-polarized light, respectively. MS2 and MS3, on the other hand, show strong dependence on $\delta$, while maximum diffraction contrast occurs for polarization along one of the respective basis vectors; that is, along $(\hat{\mathbf{a}},\hat{\mathbf{b}})$ and $(\hat{\mathbf{r}},\hat{\mathbf{l}})$ for MS2 and MS3, respectively. In order to better see that the three diffraction contrasts indeed represent normalized Stokes parameters, Fig. \ref{fig:theory}(d) shows the three-dimensional surface covered by the diffraction contrasts. As expected, the surface closely represents the Poincar\'e sphere, with the deviation from the ideal unit sphere displayed in Fig. \ref{fig:theory}(e). It is seen that the error depends on the exact polarization state but, in general, the deviation is of the order $10^{-2}$. The largest deviation from any of the Stokes parameters is found to be $4\cdot10^{-2}$, while the (2-norm) error in the vector $(D_1,D_2,D_3)$ averaged over all polarization states is $2\cdot10^{-2}$. We interpret this small error as a sign of valid approximations. Moreover, we would also like to stress that the diffraction contrast, cf. \eqref{eq:D}, is a robust parameter in the sense that it is a relative measurement and, hence, "self-calibrating". For example, deviations from the ideal reflection coefficient in \eqref{eq:r}, like position-dependent reflection amplitude [Fig. \ref{fig:design}(e)] or non-linear phase gradient, may lead to diffraction into other orders, but such a feature does not necessary affect the diffraction contrast, only the level of intensities. As an example, \href{link}{Supplement 1}, Fig. S1, shows diffraction efficiencies of the three metasurfaces as a function of wavelength in the interval 600-1100\,nm. Despite the quick deterioration of the blazed grating functionality as one moves away from the design wavelength, the first order diffraction contrasts remain high, hereby implying that designed metasurfaces can operate in a broad wavelength range. For example, if we accept a maximum deviation of $\sim0.1$ between any Stokes parameter and the associated diffraction contrast the operation bandwidth is $\sim 700-1000$\,nm, as demonstrated in \href{link}{Supplement 1}, Fig. S2.

\subsection{Experiments}
With the above numerical simulations illustrating the possibility to obtain the polarization-dependent Stokes parameters by simple measurement of the diffraction contrast in three phase-gradient metasurfaces, we now proceed with the experimental verification. The metasurfaces are fabricated on a silicon substrate using standard deposition techniques and electron-beam lithography, with 3\,nm Ti deposited between successive layers for adhesion purposes. 
Figs. \ref{fig:exp1}(a)-\ref{fig:exp1}(c) display representative images of the $75 \times 75$\,$\mu\mathrm{m}^2$ fabricated metasurfaces, demonstrating in all three cases reasonable correlation with the designs [Fig. \ref{fig:design}(d)], though discrepancies are also noticeable. 
%
%
For example, the elongated nanobricks in MS1 and MS2 have become too wide, while the nanobricks of MS3 are not completely identical. Nonetheless, the metasurfaces have been optically characterized by the setup shown in Fig. \ref{fig:exp1}(e). 
Here, the sample is fixed in a mount with XYZ translation and exposed by light from a tunable titanium sapphire laser with the wavelength set to 800\,nm. At the same time, the sample can be inspected by white light from a lamp. The polarization state of the incident light is controlled by a polarizer (P) and waveplate (WP2), where WP2 is a quarter-wave plate or half-wave plate. WP1 is an additional wave plate which together with P control the laser power incident on the sample. Once the polarization state is fixed, the light is weakly focused by a lens (L1) onto the metasurface with a spot diameter larger than the metasurface. The reflected light is collected by a long-working distance objective with numerical aperture 0.55 whose front focal plane (FFP) is coinciding with the metasurface. Note that in order to collect the reflected light, a beam splitter (BS2) is inserted between the objective and the metasurface. In order to compensate for the phase change between orthogonally polarized light induced by BS2 we insert an additional beam splitter (BS1) rotated by $90^{\circ}$ w.r.t. BS2 so as to compensate for this phase change and thereby preserve the polarization state of incident light after transmission through BS1 and BS2. The metasurface diffraction pattern is finally obtained by projecting the objective back focal plane (BFP) by another lens (L2) onto a CMOS camera which for sufficiently low optical power can be operated with a linear response. It should be noted that the non-ideal performance of the different optical components, particularly the broadband quarter-wave plate, and possible imperfections in the alignment implies an overall uncertainty in the state of polarization of the incident beam that is estimated to be $\sim 5$\,\%.

%
\begin{figure}[tb]
\centering
\includegraphics[width=7.5cm]{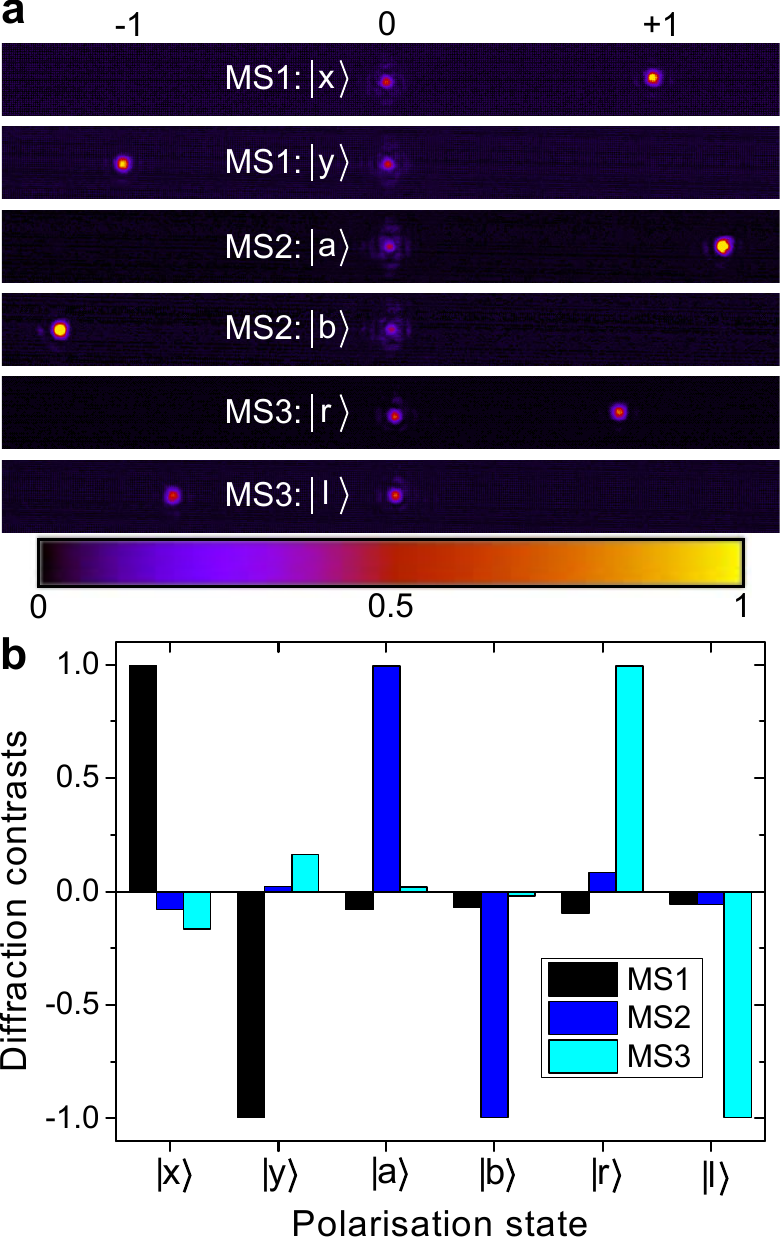}
\caption{Experimental verification of metasurface polarization-splitters. (a) Normalized optical images of the diffraction spots for the two orthogonal polarizations that the metasurfaces are designed to split. The incident polarization state is denoted by $|u\rangle$ in each panel. (b) Measured diffraction contrasts for the six polarizations that represent the extreme values of the three Stokes parameters. \label{fig:exp2}}
\end{figure}
Regarding optical characterization of the three metasurfaces, Fig. \ref{fig:exp2}(a) displays relevant images of the diffraction spots. 
It is clear that imperfections in the fabricated metasurfaces, together with uncertainties in the optical properties of evaporated gold with respect to tabular values\cite{chen2}, lead to noticeable zero-order diffraction which is, otherwise, strongly suppressed in simulations (see, e.g., \href{link}{Supplement 1}, Fig. S1). In fact, for the three metasurfaces approximately $\sim 40$\,\%, $\sim 30$\,\%, and $\sim 45$\,\% of the reflected light is contained in the zeroth diffraction order, respectively, while the remaining part of the light goes into $\pm 1$ diffraction order depending on the polarization of the incident light. That said, the three metasurfaces do demonstrate for the appropriate polarizations high diffraction contrasts, with one of the first order diffraction spots being almost completely extinguished [Fig. \ref{fig:exp2}(a)]. 
Considering the six polarization states that correspond to the extreme values (i.e., $\pm1$) of the normalized Stokes parameters, Fig. \ref{fig:exp2}(b) displays the associated diffraction contrasts measured on the three metasurfaces. It is evident that the diffraction contrasts, in general, closely represent the normalized Stokes parameters for the given polarization state $|u\rangle$, though diffraction contrasts that ideally ought to be zero in some instances show values of up to $\sim0.15$. We ascribe this deviation to the above mentioned fabrication imperfections and uncertainties in material properties with respect to the theoretical design, along with the uncertainty in the incident polarization state.

\begin{figure}[tb]
\centering
\includegraphics[width=7.6cm]{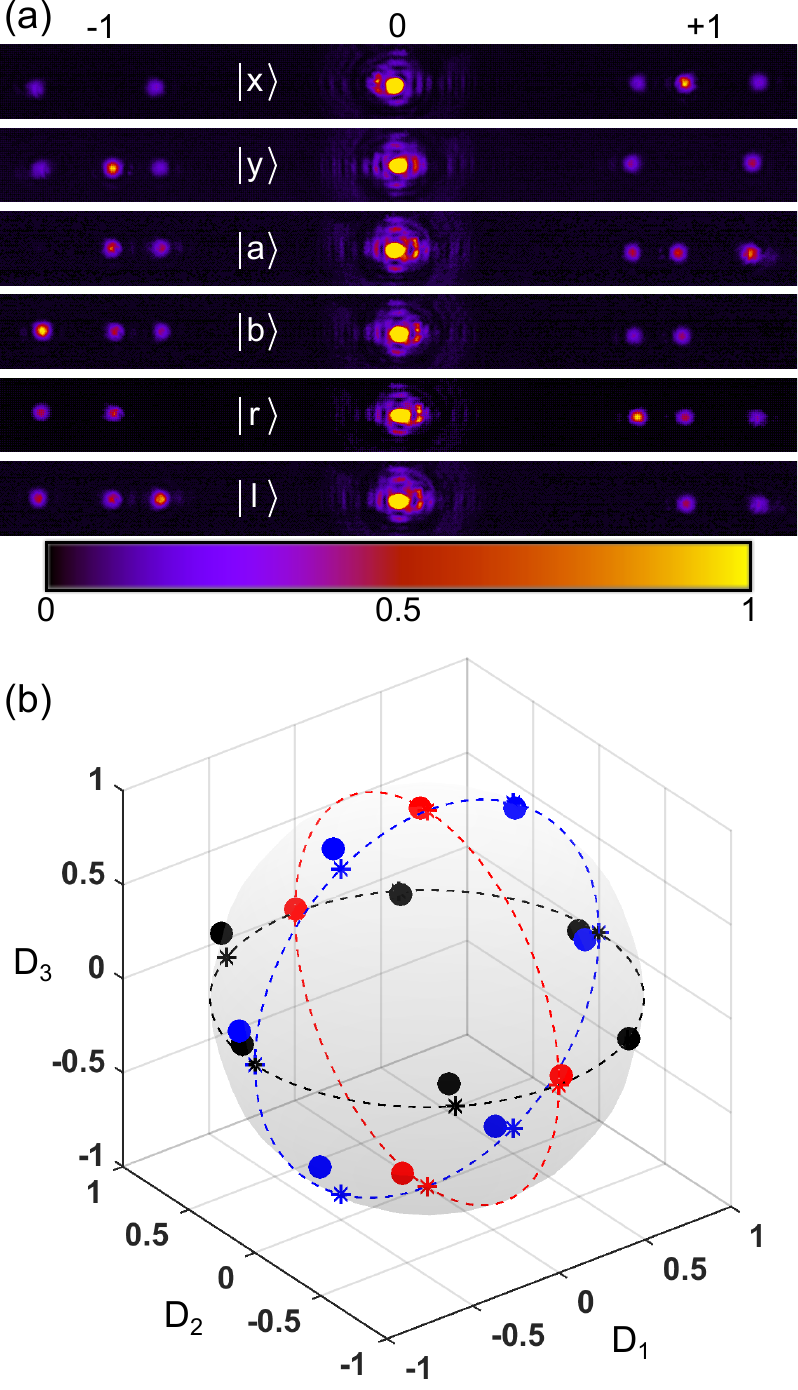}
\caption{Experimental verification of all-polarization birefringent metagrating. (a) Normalized optical images of the diffraction spots for the six polarizations representing the extreme values of the Stokes parameters. The incident polarization state is denoted by $|u\rangle$ in each panel. (b) Measured diffraction contrasts (denoted by circles) for polarization states along the main axes of the Poincar\'e sphere (indicated by asterik symbols). The measured diffraction contrasts can be found tabulated in Supplement 1, Table S1. \label{fig:exp3}}
\end{figure}

Having experimentally verified the functionality of the designed metasurfaces, we now construct an all-polarization sensitive \textit{metagrating} by interweaving the three metasurfaces. As a first thought, it seems tempting to interweave the three super cells one-by-one along the $y$-direction, as has been done in other metasurface applications\cite{nanfang_2012,shaltout_2014,ding_2015}. However, realizing that the distance between identical super cells becomes $3\Lambda_y=750$\,nm, we no longer satisfy the criteria of true sub-wavelength separation, thus significantly degrading the performance of the individual metasurfaces. As a simple solution, we could instead allow the metagrating to consist of three $\mu$m-scale blocks each representing one of the three metasurface, although such a design would require a certain size and homogeneity of the incoming beam. For the above-mentioned reasons, we realize the metagrating by interweaving the three metasurfaces in pairs of four super cells along the $y$-direction [Fig. \ref{fig:exp1}(d)], which allows for \textit{simultaneous} determination of the three Stokes parameters, while the overall functionality does not critically depend on the size and homogeneity of the incoming beam. The drawback is the introduction of unwanted first order diffraction in the $yz$-plane that contains close to $\sim 50$\,\% of the reflected light. 
That said, we would like to emphasize that the three metasurfaces constituting the metagrating work independently due to different super cell periodicities and, hence, different diffraction angles. This point is validated by considering the diffraction spots of the metagrating [Fig. \ref{fig:exp3}(a)]. 
It is clear that for the six polarizations considered, one diffraction order is in turn suppressed while the remaining two pairs of diffraction spots show (approximately) equal splitting. 
%
More importantly, considering a multitude of polarization states along the main axes of the Poincar\'e sphere, it is seen that the associated diffraction contrasts, each obtained by averaging three successive measurements, replicate reasonable well the unit sphere [Fig. \ref{fig:exp3}(b)], with points covering all octants of the three-dimensional parameter space. It should be noted that the average (2-norm) deviation between the Stokes vector of the incident light and measured diffraction contrasts is $\sim 0.1$, which is an order of magnitude larger than expected from numerical simulations. As a way to illustrate the uncertainty in the generated state of polarization of the incident beam, Fig. \ref{fig:exp3}(b) also displays two sets of diffraction contrasts for the polarization states $(s_1,s_2,s_3)/s_0=(\pm1,0,0)$ corresponding to $x$- and $y$-polarized light. Ideally, these points should pairwise be coinciding on the Poincar\'e sphere, but the fact that the linear polarizations are realized by WP2 being either a (properly oriented) quarter- or half-wave plate result in slightly different states of polarizations. To be fair, it should be mentioned that both pairs show a 2-norm deviation of $\sim 0.06$, which is also the typical deviation between successive measurements of the same state of polarization (see, e.g., \href{link}{Supplement 1}, Table S1). However, as all points shown in Fig. \ref{fig:exp3}(b) are an average of three measurements, it is more likely that the main contribution to the deviation in the two pairs relates to the uncertainty in the state of polarization. 
Finally, we remark (in accordance with simulations) that no significant degradation of the metagrating performance is observed at wavelengths 750\,nm and 850\,nm (see \href{link}{Supplement 1}, Fig. S3).

\section{Concluding remarks}
Based on phase-gradient metasurfaces in reflection, we have reported on the design and realization of three birefringent blazed gratings that split orthogonal polarizations of different bases at a wavelength of 800\,nm, while the associated diffraction contrasts equal the Stokes parameters of the incident light. The three metasurfaces are interweaved into one configuration, denoted a metagrating, that allows for the determination of the polarization state of the incident light in one measurement without the need of additional polarizers. As such, the metagrating constitutes a fast, simple, and compact way to determine a probe signal's unknown polarization state. We envision that the metagrating can be utilized as an add-on in conventional optical setups or, due to its sub-wavelength thickness, be part of an ultra-compact nanophotonic system. An interesting application might be its usage (for fast determination of the polarization state) in the feedback control for systems aimed at the synthesis of arbitrary polarization states\cite{fortuno_14}. 

It should be noted that the current experimental results show a deviation between Stokes parameters and diffraction contrast of $\sim0.1$. However, we are confident that this error can be significantly reduced with better fabrication facilities and a design that incorporates knowledge of the true optical properties of the evaporated glass and gold. Importantly, simulations and experiments have indicated a rather broadband response of the designed metasurfaces, featuring theoretical diffraction contrasts that closely represent Stokes parameters in the wavelength range $700-1000$\,nm. That said, we would like to emphasize that in the considered interval of $600-1100$\,nm the three metasurfaces demonstrate a one-to-one mapping of a certain polarization state and the resulting set of diffraction contrasts (see \href{link}{Supplement 1}, Fig. S2). For this reason, the operation bandwidth can be substantially increased if the metagrating, in advance of usage, is calibrated in the wavelength range of interest. 

Generalizing, we would like to remark that since the functionality of the metagrating is based on phase-control of the reflected light it can be extended to other wavelength ranges of interest. Also, the current configuration does not necessarily require utilization of metallic nanostructures. In fact, properly designed arrays of high-dielectric particles ought to be able of exhibiting the same functionality, with an additional benefit of potentially higher efficiencies due to negligible Ohmic losses in the particles\cite{yang_2014}. Moreover, it should be possible by utilizing the concept of Huygen's metasurfaces\cite{pfeiffer} and an all-dielectric approach\cite{lin_2014,decker_2015} to design the metagrating for operation in transmission while still preserving a high efficiency. Finally, we notice that the presented work can be regarded as complementary to a large magnitude of earlier metasurface applications, which deal with the manipulation of light (see, e.g., \cite{hao,pors_2011,khoo_2011,zhao2,roberts_2012,verslegers,lin,ishii,aieta2,chen_2012,ni,sun2}), and as such may stimulate an interest in utilizing the possibilities of metasurfaces in detection schemes.

\section*{Funding Information}
Danish Council for Independent Research (the FNU Project, Contract No. 12-124690); European Research Council, Grant 341054 (PLAQNAP).


\bigskip \noindent See \href{link}{Supplement 1} for supporting content.


\end{document}